\RequirePackage{fix-cm}
\documentclass[twocolumn,epjc3]{svjour3}  
\smartqed  
\RequirePackage{graphicx}
\journalname{Eur. Phys. J. C}
\begin{document}

\title{How to form a wormhole
}


\author{De-Chang Dai\thanksref{e1,addr1,addr4}\and Djordje Minic\thanksref{addr2} \and Dejan Stojkovic\thanksref{addr3}
}


\thankstext{e1}{e-mail: diedachung@gmail.com}


\institute{Center for Gravity and Cosmology, School of Physics Science and Technology, Yangzhou University, 180 Siwangting Road, Yangzhou City, Jiangsu Province, P.R. China 225002 \label{addr1}
           \and
           Department of Physics, Virginia Tech, Blacksburg, VA 24061, U.S.A. \label{addr2}
           \and
           HEPCOS, Department of Physics, SUNY at Buffalo, Buffalo, NY 14260-1500, U.S.A.\label{addr3}
           \and
           CERCA/Department of Physics/ISO, Case Western Reserve University, Cleveland OH 44106-7079\label{addr4}
}

\date{Received: date / Accepted: date}

\maketitle

\begin{abstract}
We provide a simple but very useful description of the process of wormhole formation. 
We place two massive objects in two parallel universes (modeled by two branes). Gravitational attraction between the objects competes with the resistance coming from the brane tension. For sufficiently strong attraction, the branes are deformed, objects touch  and a wormhole is formed. Our calculations show that more massive and compact objects are more likely to fulfill the conditions for wormhole formation.  This implies that we should be looking for wormholes either in the background of black holes and compact stars, or massive microscopic relics. Our formation mechanism applies equally well for a wormhole connecting two objects in the same universe.    

\end{abstract}

\section{Introduction}

\begin{figure}
\includegraphics[width=8cm]{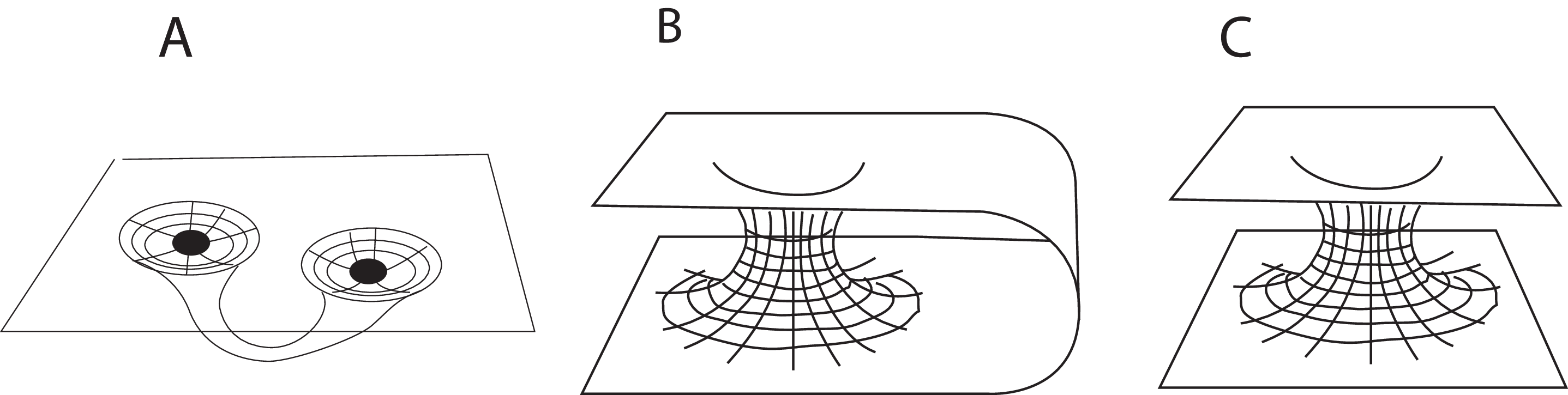}
\caption{A: A wormhole connecting two points in the same universe. B: In brane world models, the $3+1$-dimensional  universe is confined on a brane that can be twisted or folded. The wormhole structure can connect two spacetime points on brane through the bulk. C: Two completely disconnected branes are connected by a wormhole.  This and the previous option are locally equivalent.
}
\label{wormhole}
\end{figure}

Wormholes are fascinating constructs that connect two  distant spacetime points (e.g. in Fig.~\ref{wormhole} A). They are increasingly attracting attention of physicists for many different reasons \cite{wormholes} and \cite{Dai:2018vrw}.
However, so far, there is no realistic physical model of a wormhole formation. The main difficulty is the necessary presence of negative energy density  (see for example \cite{Shatskiy:2008us}), that cannot be created in macroscopic quantities. 
Quantum fluctuations can provide local negative energy density, and indeed microscopic wormholes were studied by Callan and Maldacena in \cite{Callan:1997kz}. However, the same mechanism cannot be applied for large astrophysical wormholes, which can have quite different properties \cite{Kardashev:2006nj}.

The main aim of this paper is to describe a possible mechanism of a wormhole creation under the influence of classical gravity, without invoking quantum (topology changing) effects or some other exotic physics.

There are many models in literature  in which our universe is a $3+1$-dimensional sub-space (or brane) embedded in a higher dimensional space \cite{extradim}. Such a brane, just like ordinary matter, could preserve quantum fluctuations from the epoch of its creation \cite{Saremi:2004yd,Felder:2002sv,Felder:2001kt}. These fluctuations could cause the brane to fold, twist and even cross itself. Therefore, some of the space points may be far apart along brane but indeed very close in the bulk.  

A space which is folded  (e.g. Fig.~\ref{wormhole} B), can potentially support a shortcut between two distant points. Alliteratively, a wormhole can connect two different disconnected universes (e.g. Fig.~{wormhole} C).   Locally, these two models do not differ since for the local physics it is not crucial 
to consider how these  two branes connect in the distance.  

\begin{figure}
\includegraphics[width=8cm]{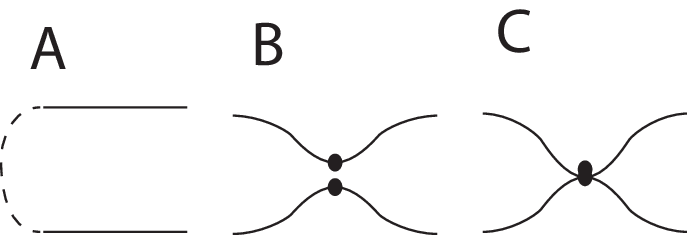}
\caption{A: Two branes represented by the solid lines are initially parallel (in the distance they could be twisted or bent). B: Massive objects placed on each brane attract each other gravitationally. The branes bend to balance the force. C: If the gravitational force overcomes the brane tensions, the contact is made and a wormhole like structure is created.    
}
\label{brane}
\end{figure}

The basic idea is represented in Fig.~\ref{brane}. We place two massive objects in two different universes  modeled by two parallel $3+1$-dimensional branes which do not intersect. The exact configuration is fully determined by the competition between two effects - the standard gravitational attraction tries to make these two objects touch, while the brane tension tries to prevent this. As a result, the branes are  bent as shown in Fig.~\ref{brane}B. However, if gravitational interaction is strong enough, the brane tension will not be sufficient to keep the objects apart, and they will touch as shown in Fig.~\ref{brane} C.  When these two branes get connected, the whole structure resembles a wormhole.  Since in the absence of a microscopic theory of spacetime we are not well equipped to describe how the two branes get smoothly sewn together,  what we aim to discuss here is the process of brane bending to the point of contact. Thus, strictly speaking our constructs are more appropriately called wormhole-like structures. However in the present text we will simply call them ``wormholes''.            

\section{A 4D brane embedded in a 5D bulk and attractive force between two masses}
\begin{figure}
\includegraphics[width=8cm]{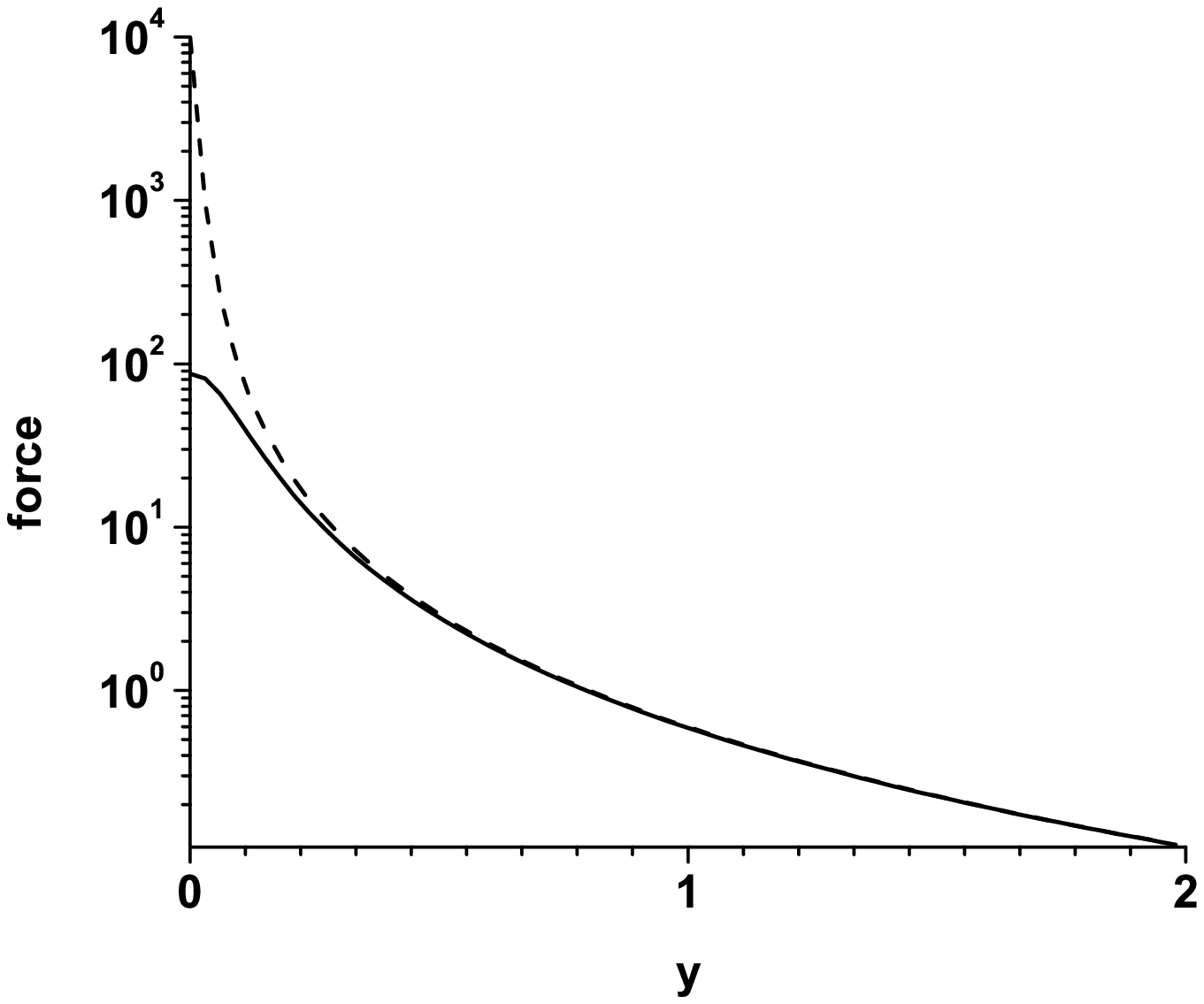}
\caption{The magnitude of the $y$ component of the attractive force between the mass $m$ on the brane and the test mass $m_t$ outside of the brane. The solid and dashed lines are for $r=0.1$ and $r=0.01$ respectively, where $r$ is the distance from the mass $m$ along the brane. For this plot we set  $\frac{1}{3\pi^2}\frac{m m_t}{M_p^2}=1$ and $r_0=1$;
}
\label{force_extra}
\end{figure}

To describe a folded space we will use a setup similar to the so-called DGP model \cite{Dvali:2000hr}. Consider a 4D space (brane) embedded in a 5D bulk.  
The Lagrangian is  
\begin{equation}
S=M^3\int R_5 dx^5 +M_p^2\int R_4 dx^4 +\int L_m dx^4 .
\end{equation}
Here, $R$ and $R_{4}$ are the 5D and 4D Ricci scalars respectively, while $L_m$ is the Lagrangian of the matter fields. In the weak field approximation the metric perturbations satisfy 
\begin{eqnarray} \label{eom}
&&\Big(M^3 \partial_A \partial_A +M_p^2 \delta(y)\partial_\mu\partial^\mu ) h_{\mu\nu} =(T_{\mu\nu} -\frac{1}{3}\eta_{\mu\nu}T^\alpha_\alpha)\delta(y)\nonumber\\
&& +M_p \delta(y)\partial_\mu\partial_\nu h_5^5 ,
\end{eqnarray}
where $M$ and $M_p$ are the 5D and 4D Planck masses respectively. $h_{\alpha\beta}$  and $T_{\alpha\beta}$ are the metric perturbations and the matter energy momentum tensor in the 4D space. $\delta(y)$ is the Dirac delta function. Capital Latin indices go over the full 5D space, Greek indices go over the 4D subspace, while $y$ is the fifth coordinate. The leading order in the gravitational potential can be obtained from $h_{00}$. The other gauge conditions can be found in \cite{Dvali:2000hr}. We can solve Eq.~(\ref{eom}) directly by the Fourier transform. The static solution is written as   
\begin{equation}
h_{00}= \int \frac{dp^3}{(2\pi)^3} \exp(i {\bf p}\cdot {\bf x}) \tilde{h}_{00}({\bf p}) \exp(-p |y|)  ,
\end{equation}
where ${\bf p}=(p_1,p_2,p_3)$ and $p=|{\bf p}|$.  

\begin{equation}
\tilde{h}_{00}=\frac{2}{3}m \frac{1}{M_p^2 p^2+2M^3p} .
\end{equation}   
Here $m$ is the mass of an object sitting in the 4D space which is the source of gravity. The gravitational potential sourced by the mass $m$ is
\begin{eqnarray}
U(r,y)&=&-\frac{1}{3\pi^2}\frac{m}{M_p^2} \frac{\Im(e^A Ei(1, A))}{r}\\
A&=&\frac{|y|-ir}{r_0}
\end{eqnarray}
where $r_0=\frac{M_p^2}{2M^3}$ is the length scale below which gravity appears as $3+1$-dimensional.
Here, $Ei(1,z)$ is the exponential integral, $Ei(1, z) = \int_1^\infty \frac{e^{-xz}}{x} dx$, while $\Im (z)$ is the imaginary part of $z$. This potential spreads both along the brane and in the bulk perpendicular to the brane. A test mass $m_t$ located outside of the brane will be attracted by the object of mass $m$ by a force in the direction perpendicular to the brane

\begin{eqnarray}
\label{force_particle}
F_5 &=& \left\{
\begin{array}{rl}-m_t\partial_y U(r,y) & \mbox{, if $r>r_s$}\\
-m_t\partial_y U(r_s,y) & \mbox{, if $r<r_s$}\end{array} 
\right.\\
\partial_y U&=&-\frac{1}{3\pi^2}\frac{m}{M_p^2} \frac{\Im \Big( \exp(A)Ei(1, A)-\frac{1}{A}\Big)  }{r_0r} ,
\end{eqnarray}
where $r$ is the distance from the mass $m$ along the brane. In the realistic case, both massive objects $m$ and $m_t$ would represent black holes or some other compact objects and therefore, they would have finite sizes.  We can treat them as point-like objects only for $r>r_s$ and $r>r_t$, where $r_s$ and  $r_t$  are the respective sizes of these objects. The magnitude of the force is shown in Fig.~\ref{force_extra}.

\section{Bending and resistance due to the brane tension}

\begin{figure}
\includegraphics[width=8cm]{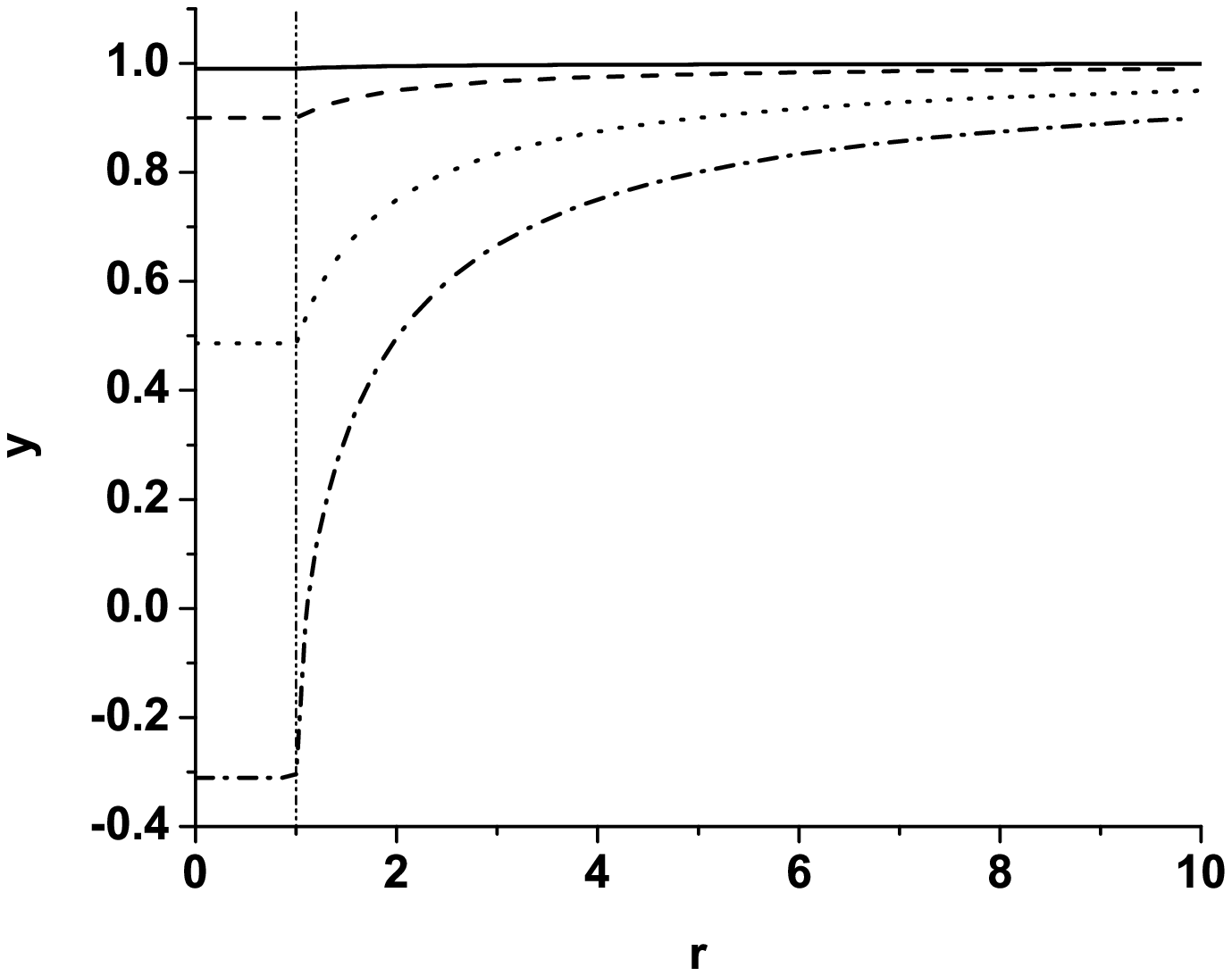}
\caption{A test object of mass $m_t$ and radius $r_t$ is misplaced from its original position $y_b=1$ to the final position $y_t$ determined by the parameter $q$, bending the brane along the way. The solid, dashed, doted and dash-doted lines represent the brane locations for $q=0.01, 0.1, 0.5, 1$ respectively. The vertical line represents the radius of the test object, that we set $r_t=1$ for the plot.}
\label{brane_bend}
\end{figure}

In the weak field approximation, the geometry in the bulk can be approximated with a flat space  
\begin{equation}
ds^2 = -dt^2 +dy^2 +dr^2+r^2 d\Omega ,
\end{equation}
where $y$ is the extra dimension, while the other four comprise the spacetime on the brane. To find the shape of the brane during the process of bending we will describe the brane with the Dirac-Nambu-Goto action. To the leading order, the brane action is  
\begin{equation}
S=-\sigma \int d^{4+1}\zeta\sqrt{\gamma} ,
\end{equation}
where $\sigma$ is the brane tension, and $\gamma$ is the determinant of the metric on the brane. 
The induced metric on the brane can be found from 
\begin{equation}
\gamma_{\alpha \beta}d\zeta^\alpha d\zeta^\beta = -dt^2 +dy^2 +dr^2+r^2 d\Omega ,
\end{equation}
where $\zeta^\alpha$ are coordinates on the brane.
The Dirac-Nambu-Goto action can be now simplified to 
\begin{equation}
S=-4\pi\sigma \int dtdr \sqrt{1+y'^2}r^2 ,
\end{equation}
where $y'=\partial_r y$. The equation of motion of the brane is 
\begin{equation}\label{yp}
\frac{y' r^2}{\sqrt{1+y'^2}}=q
\end{equation}
where $q$ is an integration constant. We will see soon that this constant is related to the bulk direction of the brane tension. The solution gives the shape of the brane $y(r)$   
\begin{eqnarray}\label{y}
&&y=y_t + \int_{r_t} \frac{q}{\sqrt{r^4-q^2}}dr=y_t+\nonumber\\
&&\sqrt{|q|} \Big(\mbox{EllipticF}( \frac{r}{\sqrt{|q|}}i,i)-\mbox{EllipticF}(\frac{r_t}{\sqrt{|q|}}i, i)\Big) ,
\end{eqnarray}
where $\mbox{EllipticF}$ is the incomplete Elliptic integral of the first kind. The equation is valid only if $r_t>\sqrt{|q|}$.  
The choice of the boundary conditions reflects the situation of interest. 
The undisturbed  brane is initially at $y_b$, i.e. $\lim_{r\rightarrow \infty}y=y_b$. We place the test object of mass $m_t$ and radius $r_t$ at the location $y_t$.   Fig.~\ref{brane_bend} shows the brane shape as  a function of $q$. As the value of $q$ grows, the brane is bent more.  This configuration describes a mass $m_t$ misplaced from its original position $y_b$ to the final position $y_t$, bending the brane along the way.

The resistance force that the brane exerts on the object $m_t$ directly depends on the  angle $\theta$ at which the brane bends with respect to the object's surface.  This angle  can be estimated from Fig.~\ref{brane_bend} as 
\begin{equation}
\cos(\theta)=dy/\sqrt{dy^2 +dr^2} .
\end{equation}
If we substitute this relation into Eq.~(\ref{yp}), we get $r^2 \cos \theta = q$.
Thus, the resistance force is
\begin{equation}
\label{force_brane}
F=4\pi \sigma r^2 \cos\theta=4\pi \sigma q\le 4\pi \sigma r_t^2 ,
\end{equation}

In Fig. \ref{force_brane} we plot the force  $F$ as a function of the brane displacement from its original position, i.e. $|y_b-y_t|$. Basically, we plot $F(q)$, where $q(y_b,y_t)$ is given by Eqs.~(\ref{yp}) and (\ref{y}).  The force reaches its maximum for $|y_b-y_t|=1.31 r_s$ and cannot grow any further with displacement. This implies that if an external gravitational force can deform the brane beyond this value, then the tension cannot counterbalance gravity anymore.  
 
\begin{figure}
\includegraphics[width=8cm]{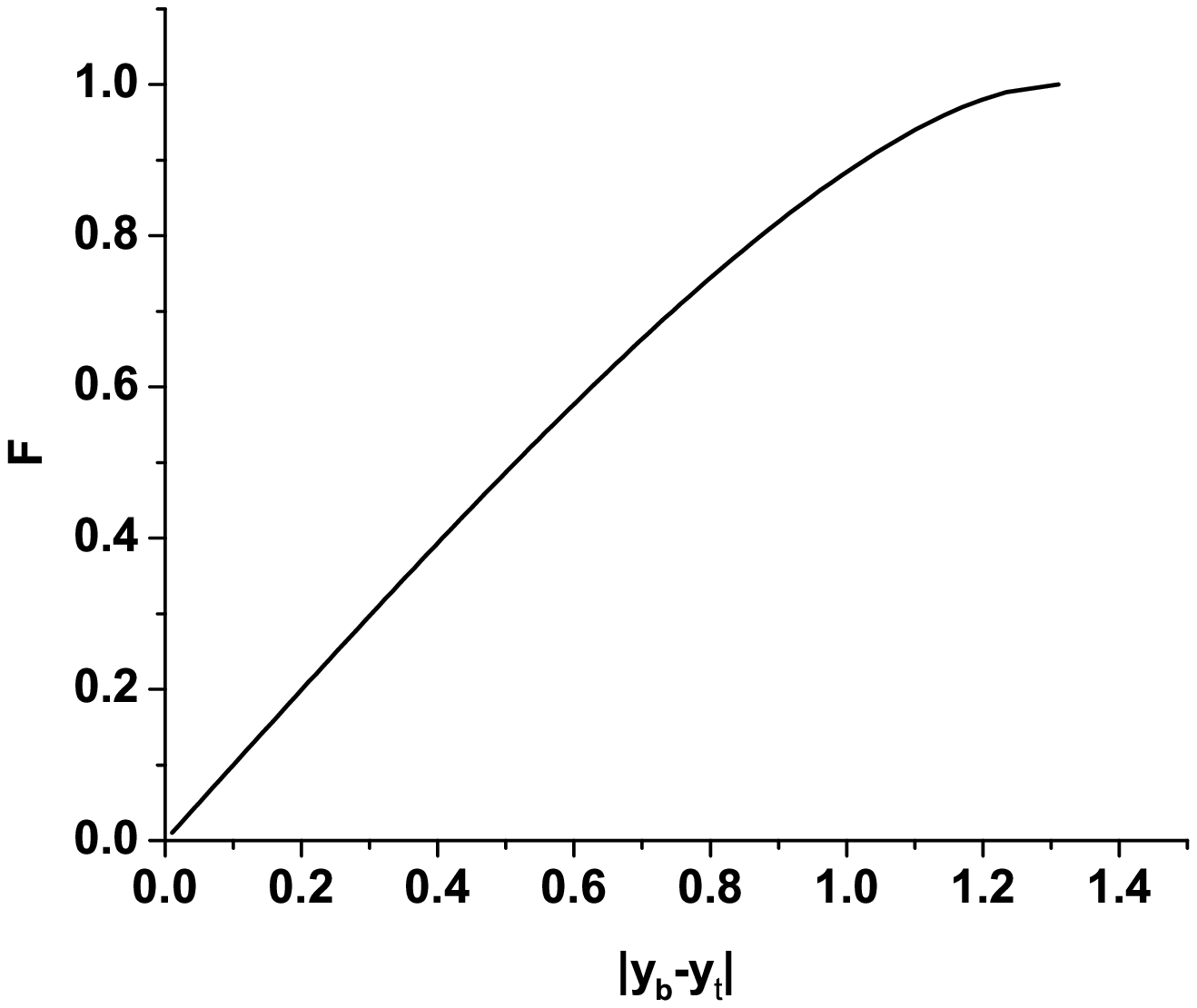}
\caption{The brane resistance force $F$ as a function of the brane displacement from its original position, i.e. $|y_b-y_t|$. The relation between brane pull force force and displacement. The unit of x-axis is  $r_s$ and  the unit of y-axis is $4\pi \sigma  r_s^2$. The maximum force appears at $|y_b-y_t|=1.31$. If the external force is large than 1, the brane tension is not enough to balance the force.   
}
\label{force_brane}
\end{figure}

\section{Balance and touching }

If the brane tension can overcome the attractive force, then the brane will bend and achieve some equilibrium. If the attractive force dominates, then these two masses will pull the branes with them and make a contact. A wormhole-like structure will be formed. Therefore,  we have to compare the two forces from Eq.~\ref{force_particle} and \ref{force_brane}. The condition is 
 
\begin{eqnarray}
\label{force_condition}
&& 4\pi \sigma q\le 4\pi \sigma r_t^2 \le  \left|\frac{m m_t\Im \Big( \exp(A)Ei(1, A)-\frac{1}{A}\Big)  }{3\pi^2M_p^2r_0r}\right|\\
&&
\end{eqnarray}

From Eq.~(\ref{force_condition}), we see that when the mass of an object gets larger, or its radius gets smaller, it is easier to get attracted to the other brane. Thus, massive and/or compact objects are more likely to produce a wormhole. 

\begin{figure}
\includegraphics[width=8cm]{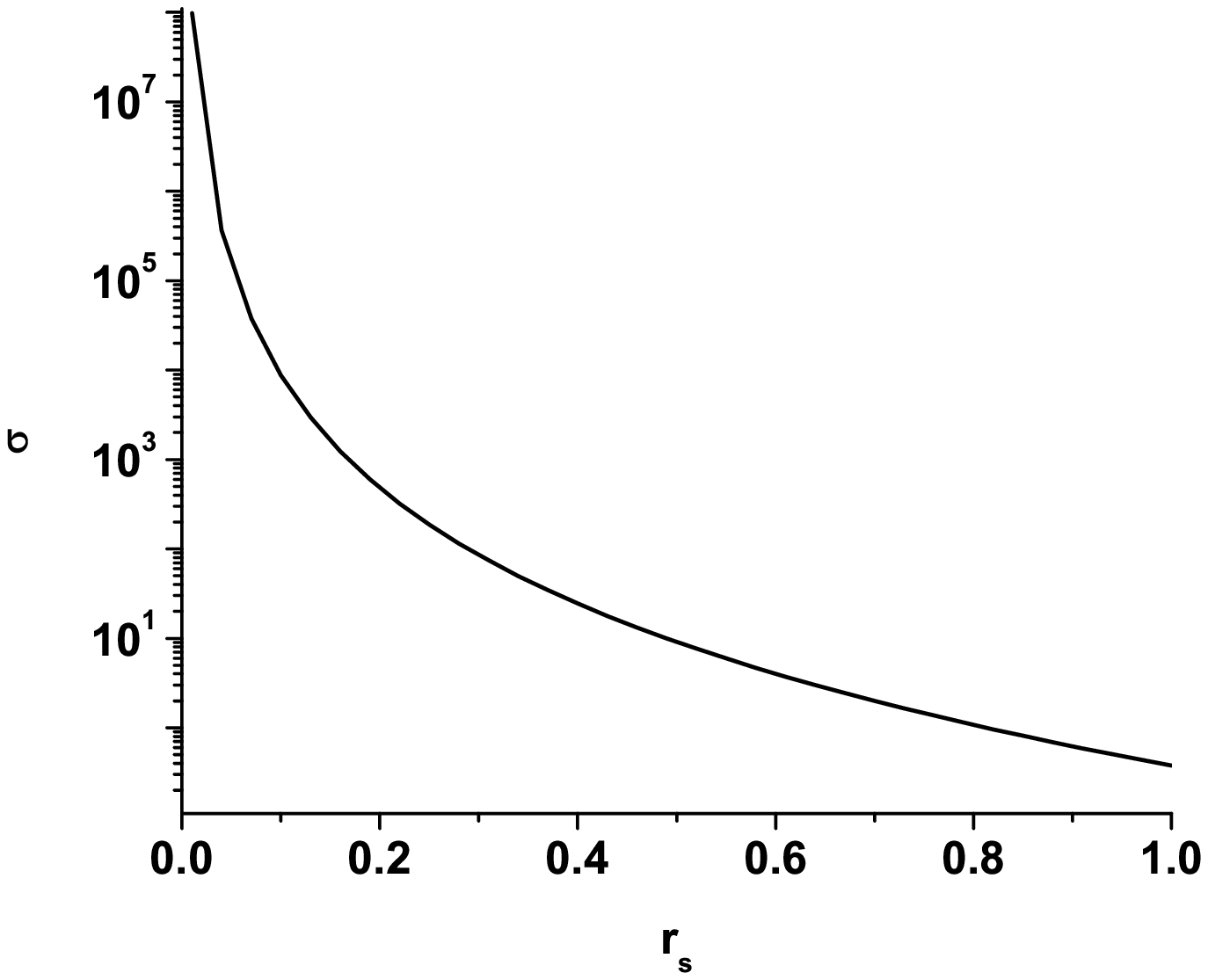}
\caption{Balance between the gravitational attraction between objects $m$ and $m_t$ and resistance due to the brane tension.  In the region below the curve, brane tension is insufficient  to prevent the massive objects (and their branes) from touching each other, and a wormhole is formed.  For this plot we set $\frac{1}{12\pi^3} \frac{m m_t}{M_p^2}=1$ and $r_s=r_t$. 
}
\label{balance}
\end{figure}

Finally, to see whether the test object with mass $m_t$ and radius $r_t$ can touch the original object of mass $m$ and radius $r_s$ we set $y=0$ and $r=r_s$ in Eq.~(\ref{force_condition}), and plot the inequality in  Fig.~\ref{balance}. This  figure shows that if the radius of the object is smaller, the tension must be bigger to prevent the object from touching the other brane. In the region below the curve, brane tension is insufficient  to prevent the massive objects and their branes from touching each other, and a wormhole is formed.

For more general calculations, one could consider a general brane location ($y_b\ne 0$). However the present  calculations are sufficient to  show that it is possible to form a wormhole that connects two initially disconnected regions
solely under the influence of gravity.

\begin{figure}
\includegraphics[width=8cm]{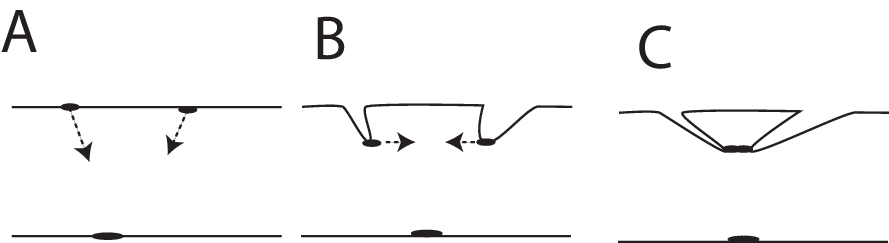}
\caption{ A: Two objects at the same brane are attracted by a much more  massive one on the other brane. B: The brane is deformed and masses are closer to each other. C: The two masses touch due to their own mutual gravity and a wormhole like structure is formed on the same brane.  
}
\label{brane_1}
\end{figure}

We note another possibility to form a wormhole in the similar framework. Consider for example a process like Fig.~\ref{brane_1}. There are two massive objects on the same brane, and one (much more massive) on the other brane that attracts them in an asymmetric way, as shown in  Fig.~\ref{brane_1} A. The brane is deformed and objects come closer to each other (Fig.~\ref{brane_1} B). Eventually the objects  touch and  a wormhole-like structure is created on the same brane (Fig.~\ref{brane_1}C).

\section{Numerical estimates}

\begin{figure}
\includegraphics[width=8cm]{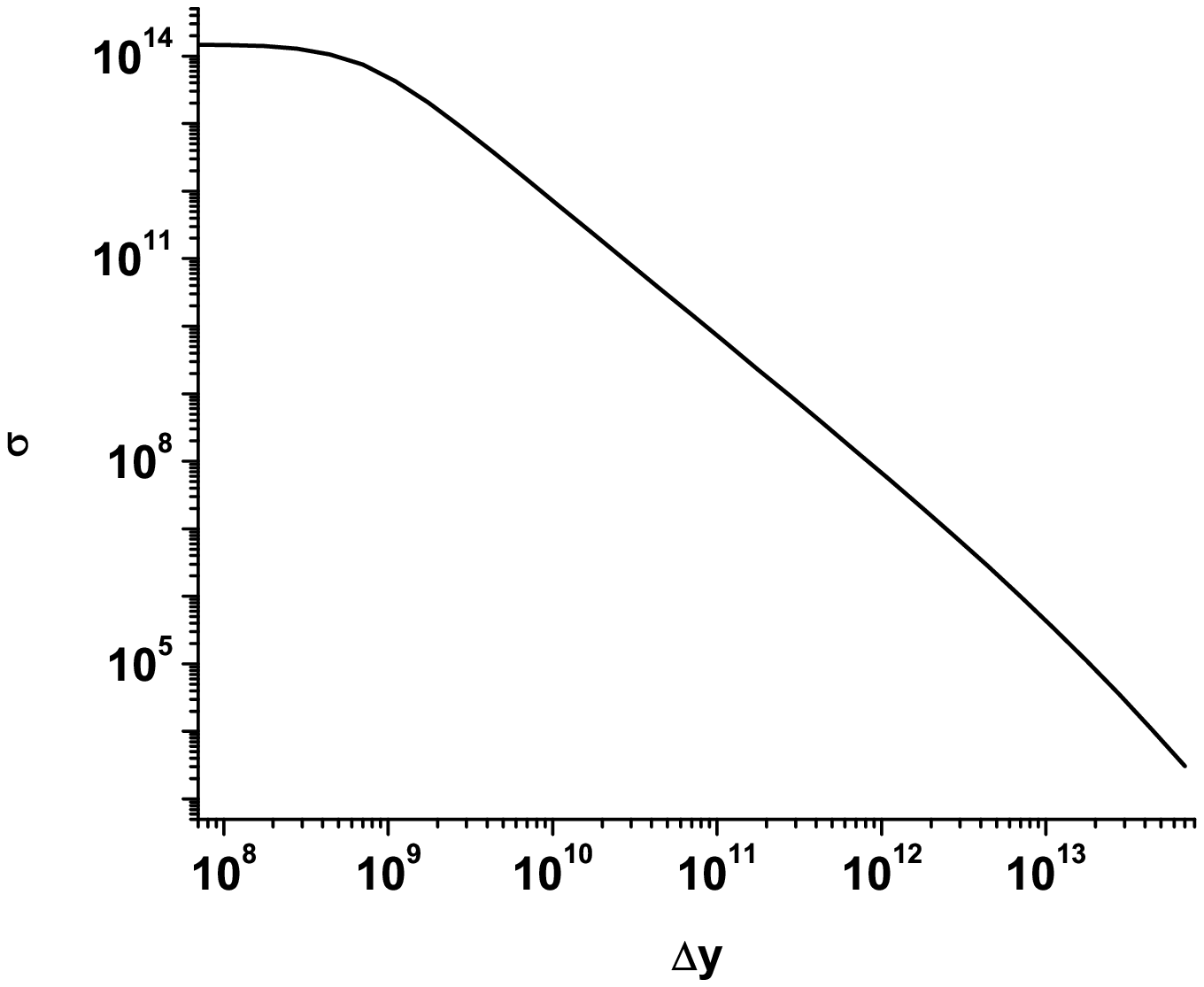}
\caption{Two solar mass stars on two different branes attracting each other. The plots shows minimal tension that can balance gravitational attraction in the direction of extra dimension. $\Delta y$ is the distance between these two branes in units of meters. $\sigma$ is the brane tension in units of $N/m^2$. Below the curve, the brane tension is insufficient to prevent the stars from touching and forming a wormhole. The other parameters are shown in the main text.   
}
\label{tension_distance}
\end{figure}

Finally, it would be very useful to estimate some numbers in the relevant parameter space. 
In particular, we are interested in realistic astrophysical objects.  
Consider for example  two solar mass stars, i.e. $m_t=m_s=M_\odot$, located on two parallel branes attracting each other. The stars' radii are $r_t=r_s\approx 7\times 10^8$m. In the DGP model that we used here, physics is $3+1$ dimensional (i.e. gravitational potential has the usual $1/r$ behavior) up to distances of  $r_0 \approx 10^{13}$m, and the Planck scale has its usual form  $M_P^2 =\frac{1}{16\pi G} $, where $G$ is the Newton's constant \cite{Dvali:2000hr}. 
Fig. \ref{tension_distance} shows the minimal tension required to balance 
 gravitational attraction in the extra dimensional direction.
Since the minimal acceptable distance between the branes is of the order of micrometers 
\cite{Lee:2020zjt}, the parameter space that allows for astrophysical wormholes is not very restrictive. 

\section{Concluding comments}

In this paper we have discussed a simple but very useful description of the process of wormhole-like formation.  Essentially, if we place two massive objects in two parallel universes, modeled by two branes, the gravitational attraction between the objects competes with the resistance coming from the brane tension, and for sufficiently strong attraction, the branes are deformed, objects touch,  and a wormhole-like configuration is formed as an end product. Our analysis indicates that more massive and compact objects are more likely to fulfill the conditions for such wormhole-like formation, which implies that we should be looking for realistic wormholes either in the background of black holes and compact stars, or massive microscopic relics. To get some feeling for the orders of magnitude, we calculated that two solar mass objects can form a wormhole like structure for reasonable values of brane tension and distance between the branes. 

Strictly speaking, what we discuss here are wormhole-like structures rather than wormholes in strict sense. We make this distinction because we deal with some global properties of the space-time rather than local geometry. The precise metric of the wormhole-like structure would depend  on the concrete massive objects we are talking about. If objects located in two parallel universes exerting gravitational force on each other are black holes, then the resulting wormhole would not be traversable due to the presence of the horizon. If the objects in question are neutron stars of other horizon-less objects, then the wormhole would be traversable. 

It is also importrant to note that the role of negative energy density which provides repulsion that counteracts gravity is played by the brane tension.  Thus we do not need extra source of negative energy density in our setup to support gravity. However, this still does not guaranty stability of the whole construct. It could happen that a very long wormhole  throat breaks into smaller pieces in order to minimize its energy. To verify this, a full stability analysis would be required. We leave this question for further investigation.  

There is a huge range in parameter space that allows for wormhole creation in our setup. Since the balance between the brane tension and mass (and size) of the objects is required, from Eq.~(\ref{force_condition}) we see that if the brane tension is zero, any non-zero mass would be sufficient to form a wormhole.  Similarly, if the brane tension is infinite, one would need an infinitely massive object to form a wormhole. Thus, apriory there is no minimal nor maximal max required to form a wormhole. However, from the plot in Fig.~\ref{balance} we see that for a fixed brane tension and fixed mass more compact objects are more likely to form wormholes.

Related issues reserved for future investigation that could possibly be answered in the same or similar framework include the question whether wormholes are produced before or after (brane of bulk) inflation, and whether they are stable on cosmological timescales.
In trying to address such difficult phenomenological issues (so far considered in a different context in \cite{Kardashev:2006nj})
one could follow the logic used in description of the production of
cosmic strings in brane world models (a good review of this topic is \cite{Polchinski:2004ia}).

Note also that cosmic strings were proposed as sources of very particular gravitational wave pulses \cite{Polchinski:2004ia} and thus one could also envision that ``snapping'' wormholes would produce very characteristic gravitational wave signals as well.
In order to answer such realistic astrophysical question we probably need to resort to  detailed numerical simulations.

Finally,  we have recently discussed how to observe wormholes \cite{Dai:2019mse}, \cite{Simonetti:2020vhw}, by studding the motion of objects in vicinity of a wormhole candidate. The same strategy can be applied in the current context. 

\begin{acknowledgements}
We thank M. Kavic and J. Simonetti for discussions. D.C Dai is supported by the National Natural Science Foundation of China  (Grant No. 11775140). D. M. is supported in part by the US Department of Energy (under grant DE-SC0020262) and by the Julian Schwinger Foundation. D.S. is partially supported by the US National Science Foundation, under Grants No. PHY-1820738 and PHY-2014021.
\end{acknowledgements}

\end{document}